\documentclass{article}

\usepackage{arxiv}

\usepackage[utf8]{inputenc} % allow utf-8 input
\usepackage[T1]{fontenc}    % use 8-bit T1 fonts
\usepackage{hyperref}       % hyperlinks
\usepackage{url}            % simple URL typesetting
\usepackage{booktabs}       % professional-quality tables
\usepackage{amsfonts}       % blackboard math symbols
\usepackage{nicefrac}       % compact symbols for 1/2, etc.
\usepackage{microtype}      % microtypography
\usepackage{lipsum}		% Can be removed after putting your text content
\usepackage{graphicx}
\usepackage{natbib}
\usepackage{doi}
\usepackage{amsmath}

\title{Metadata-Driven Retrieval-Augmented Generation for Financial Question Answering}

\author{ \href{https://orcid.org/0009-0006-3807-7216}{\includegraphics[scale=0.06]{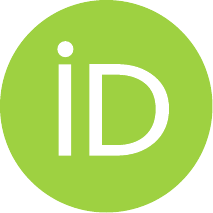}\hspace{1mm}Michail Dadopoulos} \\
	Aristotle University of Thessaloniki\\
	\texttt{mdadopoul@ece.auth.gr} \\
	\And
	{\hspace{1mm}Anestis Ladas} \\
	University of Macedonia\\
	\texttt{aladas@uom.edu.gr} \\
    \And
	\href{https://orcid.org/0000-0002-2782-4582}{\includegraphics[scale=0.06]{orcid.pdf}\hspace{1mm}Stratos Moschidis} \\
	University of Macedonia\\
	\texttt{smos@uom.edu.gr} \\
    \And
	\href{https://orcid.org/0009-0002-0899-0996}{\includegraphics[scale=0.06]{orcid.pdf}\hspace{1mm}Ioannis Negkakis} \\
	University of Plymouth\\
	\texttt{ioannis.negkakis@plymouth.ac.uk} \\
}

\renewcommand{\shorttitle}{Metadata-Driven RAG for Financial QA}

\hypersetup{
pdftitle={Metadata-Driven Retrieval-Augmented Generation for Financial Question Answering},
pdfsubject={q-bio.NC, q-bio.QM},
pdfauthor={David S.~Hippocampus, Elias D.~Striatum},
pdfkeywords={First keyword, Second keyword, More},
}

\begin{document}
\maketitle

\begin{abstract}
	% \lipsum[1]
    Retrieval-Augmented Generation (RAG) struggles on long, structured financial filings where relevant evidence is sparse and cross-referenced. This paper presents a systematic investigation of advanced metadata-driven Retrieval-Augmented Generation (RAG) techniques, proposing and evaluating a novel, multi-stage RAG architecture that leverages LLM-generated metadata. We introduce a sophisticated indexing pipeline to create contextually rich document chunks and benchmark a spectrum of enhancements, including pre-retrieval filtering, post-retrieval reranking, and enriched embeddings, benchmarked on the FinanceBench dataset. Our results reveal that while a powerful reranker is essential for precision, the most significant performance gains come from embedding chunk metadata directly with text ("contextual chunks"). Our proposed optimal architecture combines LLM-driven pre-retrieval optimizations with these contextual embeddings to achieve superior performance. Additionally, we present a custom metadata reranker that offers a compelling, cost-effective alternative to commercial solutions, highlighting a practical trade-off between peak performance and operational efficiency. This study provides a blueprint for building robust, metadata-aware RAG systems for financial document analysis.
\end{abstract}

% keywords can be removed

\section{Introduction}
Annual reports are the cornerstone of corporate disclosure, providing essential data for investors, regulators, and analysts. Yet their complexity, hundreds of pages of dense text, tables, and footnotes, makes manual analysis both time-consuming and error-prone. Traditional information retrieval methods, such as keyword search, struggle with the semantic ambiguity and contextual dependencies present in these documents.

Large Language Models (LLMs) have opened new possibilities for financial question answering, with Retrieval-Augmented Generation (RAG) emerging as a leading paradigm for grounding AI outputs in source material. However, the effectiveness of RAG depends critically on its architecture and the optimization of its core components. Basic dense retrieval often fails to identify the most relevant passages, while even more advanced hybrid retrieval and reranking approaches still treat document chunks in isolation. This narrow view prevents them from leveraging the hierarchical structure and implicit relationships within financial reports context that is essential for synthesizing information and addressing subtle queries.

In high-stakes domains such as finance, accuracy and reliability are paramount. Yet, the growing variety of RAG techniques also introduces practical challenges: stakeholders must weigh not only accuracy but also the computational costs, latency, and complexity of different implementations. Understanding how architectural choices influence retrieval quality is therefore a pressing research need.

To address these challenges, this paper proposes a novel metadata-driven, multi-stage RAG architecture that treats documents not as flat collections of chunks but as hierarchical knowledge structures. Our methodology begins with an offline pipeline in which LLMs generate multi-level metadata: document-level summaries, key entities, and thematic clusters, followed by chunk-level enrichments such as parent cluster references, associated entities, potential question-answer pairs, and “retrieval nuggets” of implicit knowledge.

Building on this metadata-rich foundation, we design an architecture that integrates pre-retrieval intelligence, metadata-enriched embeddings, and post-retrieval refinement. Specifically, our study systematically investigates three intervention strategies:
\begin{enumerate}
    \item Pre-Retrieval Optimization: Using document-level metadata for intelligent file filtering and query rewriting, thereby narrowing the search space before retrieval begins.
    \item Post-Retrieval Refinement: Expanding search results through metadata-driven entity and cluster exploration, and applying a custom reranker that combines semantic and metadata relevance.
    \item Semantic Embedding Enrichment: Embedding chunks together with their metadata to create contextually richer vectors that better capture financial semantics.
\end{enumerate}

Our findings show that combining file filtering, query rewriting, and metadata-enriched “contextual chunks” significantly improves performance compared to both baseline RAG and other advanced retrieval configurations.

We validate our approach on the FinanceBench dataset \citep{islamFinanceBenchNewBenchmark2023}, a specialized benchmark for financial question answering, and measure reliability with the RAGChecker evaluation framework \citep{ruRAGCheckerFinegrainedFramework2024a}. The results highlight not only the gains of metadata-driven architectures but also provide insights into the trade-offs between accuracy, efficiency, and computational cost.

The remainder of this paper is organized as follows. Section 2 reviews the relevant literature. Section 3 details our methodological approach, including the indexing pipeline and experimental architectures. Section 4 presents results and analysis, while Section 5 concludes with key takeaways. Section 6 lists the full set of bibliographic references.

\section{Related work}
The task of retrieving information and answering questions from annual financial documents has traditionally relied on keyword-based search methods and rule-based systems. Early approaches, such as TF-IDF \citep{saltonSPECIFICATIONTERMVALUES1973,saltonVectorSpaceModel1975} and BM25 \citep{robertsonOkapiTREC31995}, were widely used for ranking document relevance. TF-IDF evaluates the importance of a term in a document by balancing its frequency within the document against its prevalence in the entire corpus. It is computed as:

\begin{equation}
\text{TF-IDF}(t, d) = \text{TF}(t, d) \times \text{IDF}(t)
\label{eq:tfidf}
\end{equation}

where $t$ is a term (word), $d$ is a document, and:

\begin{equation}
\text{TF}(t, d) = \frac{f(t, d)}{\sum_{t' \in d} f(t', d)} 
\quad \text{and} \quad 
\text{IDF}(t) = \log \left( \frac{N}{n_t + 1} \right)
\end{equation}

are the Term Frequency (TF) and Inverse Document Frequency (IDF), with $f(t,d)$ defining the raw frequency of term $t$ in document $d$, N the total number of documents in the corpus and $n_t$ is the number of documents containing the term $t$. While TF-IDF effectively ranks documents based on term relevance, it does not account for document length variations or term saturation.

Expanding TF-IDF, BM25 introduces length normalization and a saturation mechanism, refining document ranking. The BM25 score for a document $d$ given a query $q$ is:

\begin{equation}
\text{BM25}(q, d) =
\sum_{t \in q} \text{IDF}'(t) \cdot
\frac{f(t, d) \cdot (k_1 + 1)}
{f(t, d) + k_1 \cdot \left( 1 - b + b \cdot \frac{|d|}{\text{avgdl}} \right)}
\label{eq:bm25}
\end{equation}

where:

\begin{equation}
\text{IDF}'(t) = \log \left( \frac{N - n_t + 0.5}{n_t + 0.5} + 1 \right)
\end{equation}

$|d|$ is the length of the document (number of terms), $\text{avgdl}$ is the average document length in the corpus, 
$k_1$ is a hyperparameter that controls term saturation, and $b$ is a hyperparameter that controls document length normalization.

These techniques performed well for simple, keyword-matching queries but struggled with the complexity and contextual dependencies inherent in unstructured narratives and tabular data found in annual reports. 

To address the limitations of pure keyword search, classical NLP techniques were applied to enhance text retrieval and analysis. Methods such as Named Entity Recognition (NER), part-of-speech (POS) tagging, and dependency parsing \citep{nivreEfficientAlgorithmProjective2003,tjongkimsangIntroductionCoNLL2003Shared2003} enabled more structured extraction of financial concepts, such as company names, risk factors, and monetary values. Additionally, topic modeling techniques like Latent Dirichlet Allocation (LDA) \citep{bleiLatentDirichletAllocation2003} helped categorize key themes within financial disclosures. However, these often failed to generalize across different document formats and writing styles.

The introduction of distributed word representations, such as Word2Vec \citep{mikolovEfficientEstimationWord2013} and GloVe \citep{penningtonGloVeGlobalVectors2014}, further advanced text understanding by capturing semantic relationships between words. Unlike earlier approaches that treated words as discrete symbols, these models represented words as dense vectors in a continuous space, enabling better generalization across varying contexts. These embeddings provided a foundation for neural network-based models, enhancing document retrieval and classification.

Building on these advancements, machine learning-based techniques improved information retrieval and question-answering capabilities by leveraging supervised learning. Methods such as Bi-Directional Attention Flow (BiDAF) \citep{seoBidirectionalAttentionFlow2016} and Document Reader Question Answering (DrQA) \citep{chenReadingWikipediaAnswer2017}, enabled question answering by not only retrieving relevant information but also extracting and providing direct answers from text. However, these approaches lacked the ability to generalize effectively across domains, especially when encountering specialized terminology or long documents, such as financial disclosures. The limitations of these traditional methods underscored the need for more advanced techniques capable of handling the nuanced and diverse nature of annual reports.

The next major leap came with the Transformer architecture \citep{vaswaniAttentionAllYou2017}. Transformers relied on self-attention mechanisms, allowing them to process entire sequences in parallel rather than sequentially. This advancement drastically improved efficiency and performance on long documents, making Transformers particularly well-suited for analyzing complex texts. This set the stage for large language models (LLMs), such as GPT, BERT, Llama, and Gemini \citep{brownLanguageModelsAre2020b,devlinBERTPretrainingDeep2019b,touvronLLaMAOpenEfficient2023,team2023gemini}, which demonstrated remarkable capabilities in understanding and generating human-like text, achieving state-of-the-art performance across a diverse range of language tasks. 

In parallel, researchers began adapting Transformer-based models for high-performance semantic search. While BERT-based models produced rich contextual embeddings, they weren’t optimized for similarity-based retrieval. Sentence-BERT (SBERT) \citep{reimersSentenceBERTSentenceEmbeddings2019} addressed this gap using a siamese network to generate embeddings suited for efficient semantic comparison and Dense Passage Retriever (DPR) \citep{karpukhinDensePassageRetrieval2020} adopted a bi-encoder trained on question–answer pairs to map queries and passages into a shared vector space, vastly improving the relevance and accuracy of retrieved information. Contriever demonstrated that large-scale unsupervised contrastive pre-training can rival supervised methods \citep{izacardUnsupervisedDenseInformation2021}.

Building on these advancements, further innovations emerged. ColBERT introduced a late-interaction architecture that decouples offline document encoding from lightweight, token-level query-document matching at inference time, enabling large-scale semantic search and ColBERT v2 improved both quality and storage footprint via residual compression and denoised supervision \citep{khattabColBERTEfficientEffective2020,santhanamColBERTv2EffectiveEfficient2021}. ANCE pushed performance further by continually refreshing an approximate-nearest-neighbour index during training to mine increasingly hard negatives, yielding more discriminative embeddings \citep{xiongApproximateNearestNeighbor2020}. More recently, \citet{shenRetrievalAugmentedRetrievalLarge2024} showed that an LLM can even act as its own zero-shot dense retriever, tying or surpassing dedicated dual-encoder models.

Concurrent with the rise of these general-purpose models, a significant line of research focused on specializing LLMs for the financial domain to better handle its unique lexicon and context. Early attempts at adapting BERT to financial text focused on sentiment analysis. FinBERT variants fine-tuned on news \citep{araciFinBERTFinancialSentiment2019} or SEC filings \citep{yangFinBERTPretrainedLanguage2020} consistently outperformed vanilla BERT across sentiment, NER and QA tasks. Scaling up, BloombergGPT \citep{wuBloombergGPTLargeLanguage2023} introduced a 50-billion-parameter model trained on a 363 B-token blend of terminal data and general corpora, yielding large gains on proprietary and public finance benchmarks. Open-source efforts followed, such as FinGPT, which provided a data-centric pipeline and LoRA adapters for a family of FinLLMs \citep{yangFinGPTOpenSourceFinancial2023}, while Instruct-FinGPT shows that instruction-tuning further boosts zero-shot sentiment and reasoning \citep{zhangInstructFinGPTFinancialSentiment2023}. 

However, despite the power of LLMs and dense retrievers, limitations persisted. Standalone LLMs are prone to hallucinations, producing confident but incorrect answers, and lack the ability to access up-to-date or verifiable knowledge beyond their training data \citep{zhangSirensSongAI2023a}. These issues are especially problematic in high-stakes domains like finance, where factual accuracy and source grounding are essential. These shortcomings emphasized the urgent need for a framework to ground LLM outputs in reliable, external knowledge sources \citep{gaoRetrievalAugmentedGenerationLarge2023}.

To address these challenges, researchers proposed Retrieval-Augmented Generation (RAG), an architecture designed to synergistically combine the reasoning and generation capabilities of large language models with the factual grounding of external information retrievers \citep{lewisRetrievalaugmentedGenerationKnowledgeintensive2020a,guuREALMRetrievalaugmentedLanguage2020a}. The core principle of RAG is to first use a retriever to find relevant document passages from an external knowledge base and then provide this retrieved evidence as context to the LLM. By anchoring the generation process in fetched evidence, RAG directly mitigates hallucinations and ensures responses are both factually correct and current \citep{chenBenchmarkingLargeLanguage2023a,shusterRetrievalAugmentationReduces2021a}. This approach rapidly gained momentum, becoming a central paradigm for building more trustworthy and capable AI systems \citep{fanSurveyRAGMeeting2024a}.
In complex documents such as financial annual reports, RAG has shown promise in improving the accuracy and relevance of automated question-answering systems. Research has demonstrated that well-designed RAG systems can enhance investor decision-making by providing more reliable and contextually grounded responses, with high-quality retrieval components and well-structured documents playing a crucial role in performance \citep{iaroshevEvaluatingRetrievalAugmentedGeneration2024}.

The naive RAG pipeline consists of three main stages: indexing, retrieval, and generation \citep{gaoRetrievalAugmentedGenerationLarge2023}. Indexing involves preparing and organizing the knowledge base by parsing raw data from various formats, converting it into a standardized text format, chunking data into smaller segments, embedding them into a dense vector space using pre-trained encoders, and storing them in vector databases. These embeddings serve as structured representations of the textual content, allowing efficient retrieval. Given a document chunk $d_i$, the encoding function $f$ maps it into a continuous vector space:

\begin{equation}
v_i = f(d_i)
\end{equation}

where $v_i \in \mathbb{R}^d$ represents the dense vector of the document chunk in a high-dimensional space. A user query $q$ undergoes the same transformation:

\begin{equation}
v_q = f(q)
\end{equation}

Once both the document chunks and the query are embedded, retrieval is performed by computing their semantic similarity.  
Most retrieval systems rely on cosine similarity, which measures the alignment between the query and document vectors:

\begin{equation}
\text{sim}(v_q, v_i) = \frac{v_q \cdot v_i}{\|v_q\| \, \|v_i\|}
\label{eq:cosine}
\end{equation}

where $\|v\|$ denotes the Euclidean norm of the vector.  
Higher similarity scores indicate greater semantic relevance between the document chunk and the query. The retrieval mechanism selects the top-$k$ most relevant chunks $D_k$ based on these similarity scores:

\begin{equation}
D_k = \{ d_i \mid \text{sim}(v_q, v_i) \text{ is among the top } k \}
\end{equation}

Finally, in the generation phase, the query and retrieved chunks are synthesized as a prompt for an LLM to produce a contextually grounded response.

The naive RAG pipeline, while effective in demonstrating the feasibility of combining retrieval with generation, has several limitations. Basic chunking techniques often result in the loss of important contextual relationships, and simple retrieval mechanisms may return irrelevant or redundant information. Additionally, the quality of retrieved information heavily depends on the initial query formulation, if the query is vague, underspecified, or lacks key details, the retrieval component may fail to surface the most relevant information. These limitations have spurred a wave of research aimed at enhancing each component of the pipeline \citep{wangSearchingBestPractices2024}.

One critical area of improvement is query processing, which ensures that the input query is well-structured and optimized for retrieval. Techniques such as query decomposition have been developed to break down complex or multi-faceted queries into simpler sub-queries, allowing the system to retrieve relevant information for each component \citep{chanRQRAGLearningRefine2024}. Another promising approach is Hypothetical Document Embeddings (HyDE), which generates hypothetical documents based on the query to improve the retrieval of relevant content \citep{gaoPreciseZeroShotDense2023}. Query rewriting approaches, including relevance feedback, neural rewriting, and methods like query expansion and reranking, further enhance retrieval by refining ambiguous or incomplete queries to match the structure of the indexed knowledge base \citep{chuangExpandRerankRetrieve2023,maQueryRewritingRetrievalAugmented2023}. Extending this line of work, Promptagator shows that prompting large language models to synthesize task-specific queries from as few as eight annotated examples can bootstrap effective dense retrievers in few-shot settings \citep{daiPromptagatorFewshotDense2022}. 

Indexing techniques have also seen significant advancements. Advanced tools such as LlamaParse from LlamaIndex \citep{LlamaParseTransformUnstructured} and Docling from IBM \citep{DoclingDocling}  have introduced powerful parsing capabilities, enabling systems to extract meaningful semantic units from unstructured complex documents, enabling a deeper understanding of content. Traditional fixed-length chunking methods are increasingly being replaced by semantic-aware strategies that segment documents based on their intrinsic structure and content, better accommodating the complex mix of text, tables, and graphs commonly found in financial reports \citep{yepesFinancialReportChunking2024,plonkaComparativeEvaluationEffectiveness2025}. New methods such as agentic chunking, which dynamically adapts chunk sizes based on the content's importance or complexity, and chunking overlap, which ensures context continuity by overlapping adjacent chunks, have been shown to improve retrieval precision. 

Furthermore, early work on document expansion that generated synthetic queries to enrich sparse representations, boosting retrieval recall by easing vocabulary mismatch \citep{gospodinovDoc2QueryWhenLess2023,nogueiraDoc2queryDocTTTTTquery2019,nogueiraDocumentExpansionQuery2019}, laid the conceptual groundwork for today’s chunk-level augmentation techniques in RAG pipelines. Building on these ideas, recent enterprise work on question-based retrieval with atomic units demonstrates that explicitly storing the distilled questions each chunk can answer further boosts recall in domains with specialised jargon \citep{rainaQuestionBasedRetrievalUsing2024}. Similarly, Anthropic’s Contextual Retrieval dynamically prepends a short explanatory context to each chunk before embedding, ensuring that key background details are preserved within the vector representation \citep{IntroducingContextualRetrieval}.

Building on chunk-level enrichment, another set of advancements focuses on integrating broader document context and metadata to refine the indexing process. The LongRAG model integrates passage-level semantics with global document context to reduce information loss and overcome the "lost-in-the-middle" problem \citep{zhaoLongRAGDualPerspectiveRetrievalAugmented2024a}. Complementing this, indexing systems now frequently incorporate metadata such as keywords, timestamps, and document categories, which enables more targeted, filtered retrieval \citep{chiangMetadataAppearanceRetrieved2024}. This idea is extended further by corpus-level augmentation methods like tRAG, which attach vetted, domain-specific terms to each chunk’s indexed representation, ensuring that queries with rare terminology still retrieve the correct evidence \citep{leeTRAGTermlevelRetrievalAugmented2025}.

Beyond direct enrichment, recent research has also reimagined the embedding and indexing structure itself to capture more complex relationships. Memory RAG, for example, invests more computational effort during the embedding stage to generate semantically rich representations that capture deeper relationships between entities, thereby ensuring more structured retrieval and reducing hallucinations \citep{MemoryRAGSimplified}. Hierarchical approaches, such as those using multi-level indexes, offer another structural solution. By storing high-level document representations or summaries alongside detailed chunk embeddings, these systems first retrieve relevant documents and then drill down into the associated chunks, achieving a combination of low latency and high recall \citep{arivazhaganHybridHierarchicalRetrieval2023,dongMCindexingEffectiveLong2024,liuDenseHierarchicalRetrieval2021}.

More advanced structured indexing methods move beyond flat vectors entirely by leveraging graphs and recursive structures. Graph-augmented systems like GraphRAG embed a corpus-level knowledge graph, allowing the generator to reason over explicit entity links and cross-document paths \citep{hanRetrievalAugmentedGenerationGraphs2025}. Similarly, KG²RAG \citep{zhuKnowledgeGraphGuidedRetrieval2025} uses knowledge-graph neighborhoods to surface fact-level relationships missed by purely semantic methods, while KRAGEN \citep{matsumotoKRAGENKnowledgeGraphenhanced2024} has shown that graph-aware retrieval can cut hallucinations in biomedical QA by double digits. In a different structural approach, RAPTOR constructs a recursive summary tree, abstracting documents from the bottom up \citep{sarthiRAPTORRecursiveAbstractive2024}. This allows the retriever to navigate directly to the most relevant leaf or internal node, halving irrelevant hits on long documents. Finally, a distinct approach also named LongRAG utilizes a "long retriever" and a "long reader" that operate on large 4K-token windows, preserving the cohesion of evidence that spans multiple paragraphs \citep{jiangLongRAGEnhancingRetrievalAugmented2024}.

In the retrieval phase, the naive top-k retrieval strategy has evolved into more advanced methods, giving rise to Hybrid RAG \citep{sawarkarBlendedRAGImproving2024}. Hybrid retrieval combines dense retrieval (semantic understanding) and sparse retrieval (keyword-based precision) to leverage the strengths of both approaches, ensuring semantic richness and interpretability across diverse query types and datasets. The retrieval score in Hybrid RAG is typically computed as a weighted combination of dense and sparse similarity scores:

\begin{equation}
\text{Score}(q, d) = 
\lambda \cdot \text{sim}_{\text{dense}}(q, d) + 
(1 - \lambda) \cdot \text{sim}_{\text{sparse}}(q, d)
\end{equation}

Here, $\text{sim}_{\text{dense}}(q, d)$ is the similarity score from dense vector retrieval, typically computed using the cosine similarity (Equation~\ref{eq:cosine}), while $\text{sim}_{\text{sparse}}(q, d)$ represents the sparse retrieval score, commonly computed using BM25 (Equation~\ref{eq:bm25}) or TF-IDF (Equation~\ref{eq:tfidf}).  
The parameter $\lambda$ is a tunable weighting factor that balances the influence of the two methods.

Sentence-window retrieval further enhances the process by dividing documents into contextually coherent sentence segments, embedding them, and then retrieving the relevant sentences, while including the surrounding context, the sentences that come before and after the target sentence in order to maintain coherence and completeness \citep{eibichARAGOGAdvancedRAG2024}. Filtering retrieval techniques have also been introduced to refine the process by applying constraints based on keywords or metadata associated with document chunks to filter out irrelevant or low-quality chunks, prioritizing those most aligned with the query intent \citep{poliakovMultiMetaRAGImprovingRAG2024}.

Another important enhancement is reranking, which addresses the issue of noise in initial retrieval results \citep{glassRe2GRetrieveRerank2022}. Neural rankers, such as cross-encoder models, evaluate query-document pairs with attention mechanisms to prioritize the most relevant passages. Advanced methods like pairwise or listwise ranking further optimize relevance across multiple retrieved passages simultaneously \citep{caoLearningRankPairwise2007}. Recently, LLM-based rankers have emerged as a powerful alternative, leveraging the contextual understanding and generative capabilities of large language models to dynamically reorder retrieved passages based on nuanced query intent \citep{nogueiraPassageRerankingBERT2020,adeyemiZeroShotCrossLingualReranking2023,yuRankRAGUnifyingContext2024}.

Lately, the rise of more advanced language models capable of tool use, reasoning, and dynamic prompting has inspired a new class of agentic RAG techniques that let the model to adaptively orchestrate its own search and generation, evolving from earlier innovations \citep{singhAgenticRetrievalAugmentedGeneration2025}. For example, Self-Query Retrieval enables agents to convert natural-language queries into structured metadata filters, allowing more precise control over the retrieval process. The Self-Ask pattern \citep{pressMeasuringNarrowingCompositionality2022} narrows the compositionality gap by prompting the model to pose and answer its own follow-up questions, effectively planning multi-hop searches on the fly. Earlier works such as Self-RAG \citep{asaiSelfRAGLearningRetrieve2023}, which introduced self-reflection through answer critique and re-retrieval, and IRCOT \citep{trivediInterleavingRetrievalChainofThought2023}, which interleaved retrieval and reasoning across iterative steps, have influenced many agentic RAG patterns. Their core mechanisms, reflection, iteration, and retrieval-guided reasoning, now underpin more advanced systems that dynamically adapt their behavior during generation.

Building on these trends, Retrieval-Augmented Thoughts (RAT) enables LLMs to elicit context-aware verification and reasoning steps through intermediate “thought” generation, enhancing factual consistency and multi-hop coherence in long-horizon tasks \citep{wangRATRetrievalAugmented2024} . Similarly, Self-Retrieval proposes an end-to-end paradigm where a single large language model handles both retrieval and generation, eliminating the need for external retrievers while maintaining high retrieval quality through internalized semantic understanding \citep{tangSelfRetrievalEndtoEndInformation2024} .Finally, agentic systems also incorporate ideas from DRAGIN \citep{suDRAGINDynamicRetrieval2024}, which dynamically adjusts retrieval based on the model’s internal information needs, and Self-Router \citep{liRetrievalAugmentedGeneration2024a}, which lets the model choose between retrievers or processing modes depending on query length or complexity.

Evaluating the performance of RAG systems in accounting and finance requires specialized benchmarks that address the unique complexities of the domain, such as numerical reasoning, hybrid data retrieval, and long-document contexts. For assessing the core retrieval component, the broad BEIR (Benchmarking IR) suite offers a vital, domain-agnostic baseline by testing zero-shot performance across varied tasks \citep{thakurBeirHeterogenousBenchmark2021}. Within the financial domain itself, benchmarks range in complexity. FiQA, for example, provides a foundational test for basic passage retrieval and sentiment analysis. \citep{maiaWWW18OpenChallenge2018} For more demanding and realistic scenarios, datasets like FinQA are critical for testing multi-step numerical reasoning \citep{chenFinQADatasetNumerical2021}, while TAT-QA evaluates the essential ability to synthesize information from both text and tables \citep{zhuTATQAQuestionAnswering2021}. The challenge of long-context retrieval is specifically addressed by DocFinQA, which uses complete, lengthy financial filings \citep{reddyDocFinQALongContextFinancial2024}.More recent and comprehensive benchmarks like FinanceBench, FinTextQA, and FINDER push the evaluation frontier further. FinanceBench provides questions with human-annotated evidence from full reports \citep{islamFinanceBenchNewBenchmark2023}, FinTextQA focuses on long-form answers sourced from textbooks and regulatory sites \citep{chenFinTextQADatasetLongform2024}, and FINDER is explicitly designed to test RAG systems with ambiguous, expert-level queries that mimic real-world financial analysis \citep{choiFinDERFinancialDataset2025}. 

While significant strides have been made in RAG, particularly with advancements in specialized LLMs for finance and diverse retrieval strategies, there remains a significant gap in the literature regarding the systematic evaluation of advanced, metadata-driven RAG techniques for complex, high-stakes documents like annual reports. Many existing approaches treat document chunks as isolated units, failing to leverage the inherent structural and semantic relationships present within financial disclosures. To fill this void, our study undertakes a comprehensive investigation into how the deep integration of multi-level, LLM-generated metadata across the RAG pipeline can enhance retrieval accuracy and question answering and then propose a novel, multi-stage RAG architecture.

Our approach is constructed from a combination of interventions that were systematically tested and benchmarked. The core strategies we investigate, from which our proposed model is built, include:

\begin{enumerate}
    \item Pre-Retrieval Optimization: Investigating the effectiveness of intelligent file filtering and query rewriting guided by document-level metadata.
    \item Post-Retrieval Refinement: Assessing the improvements from metadata-driven expansion and custom metadata reranking techniques.
    \item Semantic Embedding Enrichment: Examining the benefits of embedding chunk metadata directly with the text for contextually richer vector representations, creating "contextual chunks".
\end{enumerate}

Our proposed architecture specifically combines pre-retrieval optimizations (file filtering and query rewriting) with the metadata enrichment of embeddings to form a uniquely effective solution. The next section describes our novel indexing pipeline and the experimental design used to build our proposed model and rigorously evaluate it against other advanced configurations on the FinanceBench dataset.

\section{Method}

This section details the experimental framework designed to systematically evaluate the impact of metadata on RAG performance for financial document analysis. We outline the dataset and evaluation metrics, the core experimental setup, the sophisticated indexing pipeline used to generate metadata, and the specific RAG architectures benchmarked in this study, along with our proposed one.
\subsection{Dataset and Evaluation Framework}

To ensure a rigorous and domain-relevant assessment of our RAG architectures, we selected FinanceBench \citep{islamFinanceBenchNewBenchmark2023} as our primary benchmark dataset and RAGChecker \citep{ruRAGCheckerFinegrainedFramework2024a} as our evaluation framework. This combination provides a challenging testbed that mirrors real-world financial analysis tasks and a sophisticated tool for quantitative, multi-faceted quality assessment.

\subsubsection{FinanceBench: A Testbed for Financial Reasoning}

FinanceBench \citep{islamFinanceBenchNewBenchmark2023} is a benchmark specifically designed to evaluate the question-answering capabilities of LLMs on complex financial documents. Its suitability for our study stems from several key characteristics:

\textbf{Corpus Composition}: The benchmark is built upon a corpus of official corporate filings, mainly 10-K filings (about three-quarters of the corpus), plus a smaller set of 10-Q, 8-K and earnings reports, from a diverse set of publicly traded companies. These documents are notoriously long, dense, and contain a mix of narrative text, numerical data, and structured tables, making them an ideal and realistic challenge for RAG systems.

\textbf{Ecologically Valid Questions}: FinanceBench contains thousands of question-answer pairs that reflect real-world information needs of financial analysts. The questions are not simple keyword-based queries, they often require:
\begin{itemize}
  \item Numerical Reasoning: Performing calculations based on values extracted from different parts of a document.
  \item Information Extraction: Pinpointing specific metrics or statements (e.g., "What was the capital expenditure in FY2022?").
  \item Logical Inference: Synthesizing information from multiple sections to answer qualitative questions (e.g., "Is the company's primary business segment capital-intensive?").
\end{itemize}

\textbf{Ground-Truth Evidence}: Crucially, each question is annotated not only with a ground-truth answer but also with the specific "evidence strings" from the source document required to answer it. This enables a precise evaluation of the retrieval component's ability to locate the correct context.

By using FinanceBench, we test our models against a benchmark that moves beyond simple fact retrieval and demands a deeper level of contextual understanding and reasoning, directly aligning with the goals of our proposed metadata-driven architectures.
For our experiments we relied on the open-source subset of FinanceBench, which contains 150 manually annotated question/answer/evidence triples drawn from 10-K and related filings. The full benchmark spans 10,231 questions, but only this 150-case sample is publicly released, enabling reproducible evaluation without access restrictions.

\subsubsection{RAGChecker: A Framework for Fine-Grained Evaluation}

To measure the performance of our RAG pipelines, we employ RAGChecker \citep{ruRAGCheckerFinegrainedFramework2024a} a fine-grained framework that diagnoses both the retrieval and generation modules. Unlike simpler metrics that provide a single score, RAGChecker offers a suite of diagnostic metrics by first decomposing the ground-truth answers and the model-generated responses into atomic, verifiable "claims”, using an LLM-based extractor, then checks entailment relations between those claims. This claim-level approach allows for a nuanced and actionable assessment.

We focus on the following key metrics provided by RAGChecker:

Overall Quality Metrics:
\begin{itemize}
  \item Precision: Measures the correctness of the generated answer. It is the proportion of claims in the model's response that are factually supported by the ground-truth answer.
  \item Recall: Measures the completeness of the generated answer. It is the proportion of claims from the ground-truth answer that are successfully covered by the model's response.
  \item F1-Score: The harmonic mean of Precision and Recall, providing a single, balanced score for overall answer quality.
\end{itemize}

Diagnostic Retriever Metrics:

\begin{itemize}
  \item Context Precision: Measures the signal-to-noise ratio of the retrieved context. It is the proportion of retrieved chunks that are actually relevant for answering the question. A low score indicates the retriever is pulling in noisy, irrelevant information.
  \item Claim Recall: Measures the retriever's ability to fetch all necessary information. It is the proportion of ground-truth claims that are covered by the retrieved chunks. A low score indicates the retriever failed to find critical pieces of evidence.
\end{itemize}

Diagnostic Generator Metrics:

\begin{itemize}
  \item Faithfulness: Measures whether the generator is inventing information. Faithfulness is calculated as the proportion of claims in the model’s answer that are directly supported by the retrieved context.
  \item Hallucination: Measures the extent to which the generator introduces information that is unsupported or contradicted by the retrieved context. It is calculated as the proportion of claims in the model’s answer that lack such support or are contradicted.
\end{itemize}

Additional generator metrics such as Context Utilization, Noise Sensitivity and self-knowledge are also available but omitted here for brevity.

By using RAGChecker, we can move beyond a simple "correct vs. incorrect" evaluation. We can pinpoint the source of errors, whether it's a failure in the retrieval step (low Claim Recall) or a failure in the generation step (low Faithfulness), providing deep insights into the specific benefits and trade-offs of each metadata-driven technique we investigate.

\subsection{Experimental Setup}

We detail the specific software, models, and services that constitute our experimental setup. All experiments were conducted using the Python programming language, with the LangChain framework serving as the primary orchestrator for defining and executing the various components of the RAG pipelines.

\subsubsection{Large Language Models (LLMs)}

Our study utilizes a suite of different LLMs, each assigned to the task for which it is best suited.

Final Answer Generation: For the final synthesis of answers from the retrieved context, we employed OpenAI's o4-mini. This model was selected for its advanced reasoning capabilities, particularly with numerical data, and its excellent balance of performance, speed, and cost, which is critical for generating precise and concise answers from dense financial reports.

Pipeline Intelligence (Filtering \& Rewriting): For intelligent, in-pipeline tasks like metadata-driven file filtering and query reformulation, we chose gpt-4.1-mini because its speed and sufficient intelligence made it an efficient solution for these relatively simple tasks.

Offline Metadata Generation: The intensive, bulk offline task of generating summaries, clusters, and chunk-level metadata was performed using Google's Gemini 2.5 Flash. This model was chosen for its large context window, high-throughput processing speed, and cost-effectiveness, making it highly suitable for the large-scale document preprocessing required by our methodology.

Evaluation Judge: The RAGChecker framework was configured to use Google's Gemini 2.5 Flash-Lite Preview-06-17 as its "judge" LLM. Furthermore, its rapid inference speed and low cost are paramount for conducting the high volume of claim-level evaluations required for a thorough and efficient assessment of each RAG architecture.

\subsubsection{Embedding and Retrieval Infrastructure}

The core infrastructure for indexing, storing, and retrieving information was built using the following components.

Embedding Model: All text chunks were converted into dense vector embeddings using OpenAI's text-embedding-3-large. This model, which generates 3072-dimensional vectors, is renowned for its state-of-the-art performance in capturing semantic nuance, making it an excellent choice for high-fidelity semantic search.

Vector Database: We utilized Qdrant as our vector database. It was selected for its high performance, scalability, and robust native support for hybrid search, allowing the efficient storage and querying of both dense vectors and sparse keyword-based vectors within a unified index. Additionally, Qdrant supports rich payload storage and advanced filtering mechanisms, enabling fine-grained retrieval based on metadata conditions, crucial for context-aware and targeted search.

Sparse Retrieval: Sparse retrieval was implemented using Qdrant's native support for BM25-based sparse vectors. This functionality, integrated with tooling from the fastembed library for tokenization, enables efficient keyword matching that complements the semantic search of dense vectors, forming the basis of our hybrid search strategies.

Reranking Model: For our advanced baseline and several other architectures, we used Cohere's rerank-v3.5 model. This powerful reranker serves as the standard mechanism for refining the initial set of retrieved candidates, re-ordering them based on deep contextual relevance to the user's query.

\subsection{Metadata-Enriched Indexing Pipeline (Offline Phase)}

\begin{figure}
	\centering
    \includegraphics[width=1.0\textwidth]{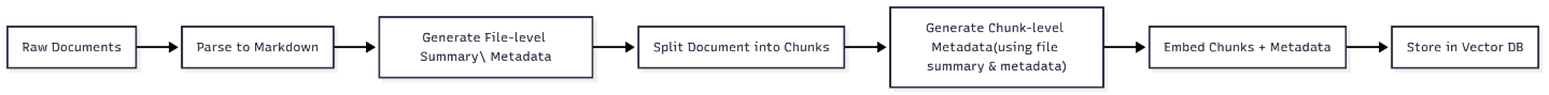}
	% \fbox{\rule[-.5cm]{4cm}{4cm} \rule[-.5cm]{4cm}{0cm}}
	\caption{Offline Indexing Pipeline}
	\label{fig:fig1}
\end{figure}

The foundation of our advanced RAG architectures is a sophisticated offline indexing pipeline designed to enrich the source documents with multiple layers of metadata. This pre-processing phase is critical for enabling the targeted retrieval strategies evaluated in this study. The entire pipeline, executed once per document, transforms raw financial reports into a structured, queryable knowledge base. The process, orchestrated via Python and LangChain, unfolds in the sequential stages detailed below and illustrated in Figure \ref{fig:fig1}.

\subsubsection{Document Parsing and Structuring}

The initial step involves parsing the raw PDF documents from the FinanceBench corpus into a clean, structured format.
PDF to Markdown Conversion: We utilize the DoclingLoader from the langchain-docling library to convert each PDF file into Markdown format. Docling is an open-source toolkit for robust document parsing and format conversion \citep{DoclingDocling}. This approach was chosen over simple text extraction as it preserves critical structural elements such as headings, tables, and lists, which provide implicit semantic context that is valuable for subsequent processing steps. The output of this stage is a single, clean Markdown text file for each original PDF.

\subsubsection{Document-Level Metadata Generation}

Once parsed, each document undergoes a high-level analysis to generate a holistic, document-wide summary. This is achieved using a structured output chain with Google's Gemini 2.5 Flash model. For each document, we generate the following metadata:

one\_liner: A single, impactful sentence that captures the document's core purpose, serving as an executive headline.

summary: A dense, analytical brief written from a third-person perspective. This is not merely a list of topics but a synthesis of the document's key findings, connecting financial figures to the strategic narrative.

clusters: A list of 5-20 high-level thematic labels (e.g., 'Financial Performance \& Results', 'Risk Factors \& Disclosures') that represent the primary topics within the document. These clusters serve as a foundational taxonomy for categorizing the document's chunks later.

This document-level metadata is stored in a JSON file, creating a master index that can be referenced by later stages of the pipeline and by the online retrieval system.

\subsubsection{Chunking and Chunk-Level Metadata Enrichment}

With document-level context established, the pipeline proceeds to the micro-level by splitting the documents and enriching each individual text chunk.

Text Splitting: Each document’s Markdown content is segmented into smaller text chunks using LangChain’s widely adopted RecursiveCharacterTextSplitter. The splitter works by repeatedly stepping down a hierarchy of separators—headings → paragraphs → sentences → words → characters, until it produces chunks no larger than the specified limit, so the text is cut at natural boundaries whenever possible. A chunk size of 1000 tokens with an overlap of 100 tokens was used to ensure semantic continuity between adjacent chunks without excessive redundancy.

Chunk-Level Metadata Generation: This is the most computationally intensive step, where each individual chunk is analyzed by the Gemini 2.5 Flash model. The model is provided with both the text of the chunk and the previously generated document-level summary and clusters as context. Through a parallelized process, it generates a rich set of metadata for each chunk:

parent\_clusters: Assigns the one or two most relevant thematic clusters from the document level to the chunk.

chunk\_entities: A list of key entities (companies, products, individuals) mentioned specifically within the chunk's text.

answered\_questions: A list of 3-10 specific, high-value questions that could be answered, either wholly or in part, by the information contained within that chunk. This proactively anticipates potential user queries.

retrieval\_nuggets: A list of non-obvious, implicit insights or connections that a sophisticated retrieval system might leverage (e.g., "Connects revenue decline to supply chain issues mentioned in the risk factors section.").

\subsubsection{Vectorization and Storage}

The final stage of the offline pipeline is to embed and store the processed data in our vector database for efficient retrieval. We create two distinct collections within our Qdrant vector store to support different retrieval strategies:

Standard Chunk Collection: The raw text of each chunk is embedded using OpenAI's text-embedding-3-large (for dense vectors) and Qdrant/bm25 (for sparse vectors) and stored. This collection serves as the baseline for traditional RAG approaches.

Contextual Chunk Collection: A second, enriched collection is created. Here, the generated chunk-level metadata (parent\_clusters, answered\_questions, etc.) is prepended to the raw chunk text before vectorization. This creates "contextual chunks" where the vector embedding itself is biased with the rich metadata, theoretically making it more semantically aligned with complex queries.

All metadata fields (file name, year, quarter, document summary, chunk entities, etc.) are stored alongside both vectors in Qdrant's payload, making them available for filtering and inspection during the online retrieval phase.

\subsection{Experimental Design - RAG Architectures Under Investigation (Online Phase)}

To comprehensively evaluate the impact of metadata on retrieval performance, we designed and tested a series of increasingly sophisticated RAG architectures. Each architecture represents a distinct strategy, building upon the last to isolate the effects of different components. All pipelines were executed against the FinanceBench dataset, retrieving the top 7 chunks per query, with the final answer generated by o4-mini based on the context retrieved by each specific architecture.

\subsubsection{Baseline Architectures}

\begin{itemize}
  \item Architecture 1: Naive RAG (Dense Retrieval). This is the simplest RAG implementation. It performs a direct dense vector similarity search against the Qdrant collection for the user's raw query.
  \item Architecture 2: Hybrid Retrieval (Combined Dense and Sparse Retrieval). This architecture enhances retrieval by combining keyword search with semantic search.
  \item Architecture 3: Hybrid Retrieval and Reranking. Building on hybrid retrieval, this architecture incorporates an additional reranking step to further refine context relevance. It first retrieves a broader set of 25 potential candidate chunks using hybrid retrieval. These are then precisely re-ordered by a powerful cross-encoder, which assesses their deep contextual relevance, ultimately providing the top 7 most relevant chunks for the final context.
\end{itemize}

\subsubsection{Pre-Retrieval Optimization Strategies}

\begin{itemize}
  \item Architecture 4: Pre-retrieval File Filtering and Query Rewriting. This is the first architecture that actively uses the pre-computed, document-level metadata, introducing pre-retrieval intelligence through filtering and query rewriting to narrow the search space and improve query precision. The retrieval process begins with file filtering where the user's query and the one-liner summaries of all documents are passed to gpt-4.1-mini. Using a structured output prompt, the LLM selects the most relevant filenames. Following this, for query rewriting, the original query, along with the full summary and clusters from the selected files, is passed to gpt-4.1-mini. The model then reformulates the query, enriching it with specific keywords and concepts to make it more effective for vector search. Finally, in filtered retrieval, the reformulated query is used to perform a standard hybrid search and reranking (as in Architecture 3), but the search is strictly filtered to only include chunks from the files selected in the first step. 
\end{itemize}

\subsubsection{Post-Retrieval Refinement Strategies}

\begin{itemize}
  \item Architecture 5: Custom Metadata Reranker. This architecture replaces the commercial Cohere reranker with our custom Metadata Reranker to test the value of explicit, feature-engineered relevance signals. The retrieval process begins by obtaining an initial set of 25 candidate chunks using one of the preceding methods. Then our reranker calculates a composite score for each chunk based on four weighted components: entity\_freq (the frequency of a chunk's entities across the entire candidate set), cluster\_coherence (the prevalence of a chunk's assigned clusters within the candidate set), entity\_query (the degree of match between a chunk's entities and terms in the user query), and retrieval (the original, normalized retrieval score from the initial search). Each component contributes equally (25\%) to the final score to ensure a balanced influence. Chunks are re-ordered based on this new composite score, and the top 7 are selected. 
  \item Architecture 6: Pre-retrieval File Filtering, Query Rewriting and Metadata-Driven Chunk Expansion. This architecture builds directly on the previous ones by adding a second retrieval step that uses chunk-level metadata to find supplementary information. The goal is to first find a few high-certainty chunks and then use their metadata to "discover" other relevant chunks that might have been missed by the initial vector search. The retrieval process first executes the full pipeline from Architecture 4 to retrieve an initial set of 4 chunks. For metadata extraction, the parent\_clusters and chunk\_entities are extracted from the metadata of these initial 4 chunks, and the most common clusters and entities are identified as "core" concepts. An expansion search is then performed as a second, metadata-only search on Qdrant, that filters for chunks sharing the "core" parent\_clusters or chunk\_entities, while excluding the chunks already retrieved, resulting in an additional 3 chunks, making the final retrieved context along the 4 initial chunks.
\end{itemize}

\subsubsection{Experiments with Contextual Embeddings}

A final set of experiments was conducted by systematically re-running previous Architectures using the Contextual Chunk Collection. In this collection, the chunk-level metadata (entities, questions, etc.) was prepended to the chunk text before being embedded by text-embedding-3-large. The hypothesis for these experiments was that by "baking" the metadata directly into the vector representation, the embeddings themselves would become more semantically aligned with complex, high-level queries, potentially improving the performance of all subsequent retrieval and reranking steps.

\section{Results and Analysis}

This section presents and analyzes the empirical results from our evaluation of the various RAG architectures on the FinanceBench dataset. The performance of each pipeline was measured using the RAGChecker framework, and the key findings are detailed below. Our analysis dissects the results from two primary perspectives: retrieval quality, assessed by Claim Recall and Context Precision, and generation quality, assessed by F1-score, Faithfulness, and the rate of Hallucination.

\subsection{Overall Performance Comparison}

The primary results for all evaluated architectures are consolidated in Table 1. The pipelines are assessed across key quality metrics from the RAGChecker framework, including overall F1-Score, Precision, and Recall, as well as diagnostic metrics for retrieval quality (Claim Recall, Context Precision) and generator performance (Faithfulness, Hallucination). At a high level, the results demonstrate a clear and progressive improvement in performance as architectural complexity and metadata integration increase.

Our central hypothesis, that a sophisticated, metadata-driven architecture can yield significant gains, is strongly validated by the results. The most successful architectures were those that integrated multiple advanced components. The peak F1-score of 44.4 was achieved by Architecture 4, which utilized Filtering, Rewriting, and a commercial Cohere Reranker on Contextual Chunks (Ctx).

Crucially, our proposed architecture, which substitutes the commercial reranker with our custom, metadata-aware model, achieved a highly competitive F1-score of 43.2. This result is a cornerstone of our findings, representing a 31\% relative improvement over the Naive RAG baseline (F1-score: 32.9) while operating faster and at zero marginal cost compared to the commercial alternative. This demonstrates that a well-designed, open-source approach can achieve performance nearly on par with black-box commercial systems, presenting a compelling trade-off for practical applications.

At the other end of the spectrum, the weakest performing model was the Hybrid Retrieval on standard chunks without a reranking step (F1-score: 30.4), underperforming even the Naive RAG baseline. This highlights a critical finding: advanced retrieval strategies like hybrid search can introduce noise and degrade performance if not paired with a robust reranking mechanism.

\begin{table}[ht]
	\caption{Comprehensive Performance of All Evaluated RAG Architectures}
	\centering
    \renewcommand{\arraystretch}{1.15}

    \resizebox{\textwidth}{!}{
    \begin{tabular}{llcccccccc}
    \toprule
    \textbf{Architecture} & \textbf{Chunk Type} & \textbf{Prec} & \textbf{Recall} & \textbf{F1-score} & \textbf{Claim Recall} & \textbf{Context Precision} & \textbf{Faith.} & \textbf{Halluc.} & \textbf{Mean Latency (s)} \\
    \midrule
    Naive RAG & Std & 35.0 & 52.7 & 32.9 & 45.9 & 20.0 & 76.4 & 18.5 & 10.47 \\
    Hybrid Retrieval & Std & 35.0 & 46.1 & 30.4 & 41.1 & 16.1 & 67.9 & 26.0 & 12.31 \\
    Hybrid Retrieval & Ctx & 38.7 & 58.4 & 37.6 & 45.1 & 17.6 & 79.1 & 14.6 & 9.26 \\
    Hybrid Retrieval and Reranking (Cohere) & Std & 37.1 & 60.2 & 38.9 & \textbf{50.7} & 23.0 & \textbf{81.7} & \textbf{12.2} & 10.7 \\
    Hybrid Retrieval and Reranking (Cohere) & Ctx & 43.0 & \textbf{70.3} & 44.1 & 48.2 & 22.3 & 77.5 & 14.3 & 10.39 \\
    Hybrid Retrieval and Reranking (Ours) & Std & 40.0 & 54.0 & 37.1 & 41.3 & 17.0 & 71.9 & 20.2 & 9.00 \\
    Hybrid Retrieval and Reranking (Ours) & Ctx & 41.5 & 62.6 & 40.6 & 43.6 & 18.9 & 74.5 & 15.7 & 9.58 \\
    Filtering + Rewriting + Hybrid Retrieval + Reranking (Cohere) & Std & 38.4 & 60.9 & 37.3 & 47.7 & 22.0 & 78.9 & 14.7 & 14.35 \\
    Filtering + Rewriting + Hybrid Retrieval + Reranking (Cohere) & Ctx & 43.8 & 69.1 & \textbf{44.4} & 42.3 & \textbf{44.4} & 76.4 & 15.1 & 13.18 \\
    Filtering + Rewriting + Hybrid Retrieval + Reranking (Ours) & Ctx & \textbf{44.1} & 64.5 & 43.2 & 39.9 & 18.0 & 73.7 & 17.2 & 12.43 \\
    Filtering + Rewriting + Hybrid Retrieval + Reranking (Cohere) + Chunk Expansion & Std & 33.6 & 57.6 & 33.2 & 41.1 & 18.3 & 71.5 & 22.2 & 13.66 \\
    Filtering + Rewriting + Hybrid Retrieval + Reranking (Cohere) + Chunk Expansion & Ctx & 40.8 & 66.2 & 40.3 & 39.5 & 18.5 & 73.9 & 18.0 & 14.87 \\
    \bottomrule
    \end{tabular}
    }
	\label{tab:rag_performance}
\end{table}

\subsection{Analysis of Retrieval and Generation Dynamics}

Here, we dissect the results to understand how each architectural choice impacted the two core stages of the RAG process: retrieval and generation.

\subsubsection{From Naive to Advanced: The Power of Reranking}

Retrieval Analysis: Moving from Naive RAG (Claim Recall: 45.9, Context Precision: 20.0) to a full Hybrid + Reranking (Cohere) pipeline (Claim Recall: 50.7, Context Precision: 23.0) shows a marked improvement in retrieval quality. The reranker was particularly effective at improving Context Precision, filtering out noisy documents retrieved by the hybrid search.
Generation Analysis: This improved retrieval context had a direct and significant impact on generation. The F1-score jumped from 32.9 to 38.9. Most notably, the Faithfulness score rose from 76.4 to 81.7, and the Hallucination rate was nearly halved, dropping from 18.5 to 12.2. This provides clear evidence that a better retrieval context directly leads to more factual and reliable generation.

\subsubsection{The Impact of Contextual Embeddings (Std vs. Ctx)}

Retrieval Analysis: The impact of using contextual chunks on retrieval metrics was not uniformly positive. While these chunks improved nearly all metrics in several configurations, they also led to a decrease in Claim Recall in the most advanced pipelines. For instance, in the "Filtering + Rewriting + Reranking (Cohere)" architecture, switching from Standard to Contextual chunks caused Claim Recall to drop from 47.7 to 42.3. This suggests that while contextual metadata helps align the chunk with the query's broader intent, it can occasionally de-emphasize the specific keywords needed to match ground-truth claims.

Generation Analysis: Despite the mixed retrieval results, using contextual chunks delivered a clear and powerful benefit to the final answer quality. In every single head-to-head comparison, using Ctx chunks resulted in a higher F1-score. For the Hybrid + Reranking (Cohere) model, the F1-score rose substantially from 38.9 to 44.1. This indicates that even if the retriever misses a few claims, the contextual information baked into the retrieved chunks helps the generator LLM to better reason and synthesize a more complete and accurate final answer. It effectively makes the generator "smarter" with the context it receives.

\subsubsection{Evaluating Pre-Retrieval Intelligence}

Retrieval Analysis: Surprisingly, adding the Filtering + Rewriting step did not consistently improve retrieval metrics. When added to the "Hybrid + Reranking (Cohere)" pipeline, Claim Recall dropped slightly from 50.7 to 47.7. This counter-intuitive result suggests that our query rewriting model, while aiming for precision, may occasionally reformulate the query in a way that misses some key terms necessary to retrieve all evidence, or the file filtering step may be too aggressive.

Generation Analysis: Despite the slightly weaker retrieval metrics, the final generation quality remained high and even saw a slight boost on contextual chunks (F1-score 44.1 to 44.4). This highlights a key trade-off: pre-retrieval steps can focus the search on highly relevant documents at the risk of missing some peripheral evidence, but the overall impact on the final answer can still be positive due to the high quality of the selected documents.

\subsubsection{Alternative Efficient Reranker Design}

Retrieval \& Generation Analysis: Our custom, metadata-aware reranker proved to be a highly successful component of the pipeline. In all experiments, it delivered a substantial performance boost over having no reranker at all. For example, on contextual chunks, adding our reranker improved the F1-score from 37.6 to 40.6.

When compared to the commercial Cohere reranker, our model's performance is slightly lower but remains highly competitive. For instance, the Cohere-powered pipeline achieved an F1-score of 44.4 on contextual chunks, while ours reached a strong 43.2. This result is very encouraging, as our reranker offers the significant advantages of being faster in every tested configuration and completely free to operate, eliminating API costs and vendor lock-in. This demonstrates that a well-designed, feature-based reranker can serve as a powerful and efficient alternative to commercial models.

\subsubsection{An Unexpected Finding: The Negative Impact of Chunk Expansion}

Retrieval Analysis: The most surprising result came from the Chunk Expansion technique. In its initial implementation (4+3 chunks), it severely degraded performance. Claim Recall for the standard pipeline plummeted from 47.7 to 41.1. This strongly suggests that the logic for finding "related" chunks based on entities and clusters was flawed, primarily adding noise rather than valuable new context.

Generation Analysis: The generation model's performance collapsed accordingly. The F1-score dropped from 37.3 to 33.2, and the Hallucination rate spiked from 14.7 to 22.2. The generator was clearly confused by the low-quality, expanded context.

\subsection{Summary of Key Findings}

\begin{enumerate}
    \item Generation Quality is a Direct Function of Retrieval Quality: Improvements in Claim Recall and Context Precision consistently led to higher F1-scores and Faithfulness, and lower Hallucination rates.
    \item Reranking is Non-Negotiable: A powerful reranking step is the single most important component for moving beyond a naive baseline. It is essential for noise reduction and improving context precision.
    \item Contextual Embeddings are a Powerful Enhancement: Enriching the content of chunks with their own metadata before embedding provides a consistent and significant performance boost across multiple architectures.
    \item Pre-retrieval Steps are a Double-Edged Sword: While aiming to improve precision, techniques like file filtering and query rewriting can inadvertently harm recall by over-constraining the search. Their value depends heavily on the quality of the controlling LLM.
    \item A Custom Reranker Presents a Viable and Efficient Alternative: While our metadata-based chunk expansion was not effective, our metadata-based reranker successfully closed most of the performance gap with a leading commercial model. Its slightly lower F1-score is coupled with superior latency and zero operational cost, presenting a practical and powerful trade-off between peak performance and deployment efficiency. This confirms that bespoke, metadata-aware components can be a cornerstone of effective and accessible RAG systems.
\end{enumerate}

\section{Conclusion, Discussion, and Future Work}

\subsection{Conclusion}

This paper sought to move beyond baseline RAG implementations and engineer a robust, metadata-aware architecture for the complex domain of accounting and financial document analysis. Through a systematic evaluation on the FinanceBench dataset, we have demonstrated that the path to high-quality, trustworthy answer generation is paved with a series of deliberate architectural enhancements.

Our investigation confirmed that a powerful reranking step is critical for precision and that enriching document chunks with their own metadata provides a significant performance boost. The central contribution of this work is the validation of our proposed multi-stage architecture, which combines LLM-driven pre-retrieval optimizations with these "contextual chunks" and a custom, transparent reranker. This architecture achieved performance nearly on par with a state-of-the-art commercial system while offering superior latency and zero operational cost. This study provides a clear blueprint for building advanced, cost-effective, and auditable analytical systems that can handle the nuance of financial information.
%For the AIS community, t
\subsection{Discussion}

The findings of this paper offer several significant contributions to the field of Accounting Information Systems (AIS), providing a practical blueprint for leveraging generative AI to navigate the complexities of financial and regulatory documents.

First, our results advocate for a metadata-first approach to designing information retrieval systems for accounting. Annual reports, SEC filings (e.g., 10-Ks), and regulatory standards are not unstructured narratives, they are highly structured documents where context is defined by sections, tables, and footnotes. Traditional methods that treat these documents as flat text lose this critical structural information. Our "contextual chunks" technique provides a method to preserve this semantic structure, effectively creating a more intelligent representation of the source document within the system. For the AIS field, this is a direct response to the challenge of analyzing complex financial disclosures. It enables the development of systems that can more accurately answer queries related to specific accounting treatments, locate relevant audit evidence, or perform automated compliance checks, as the AI has a richer understanding of not just what the text says, but where it comes from.

Second, this study presents a crucial discussion on the trade-offs between performance, cost, and auditability in AI systems for accounting. The comparison between our custom, open-source reranker and the commercial alternative speaks directly to key concerns in AIS design and governance. While proprietary, "black-box" AI models may offer marginal performance gains, our proposed architecture demonstrates that an in-house, transparent model can achieve highly competitive results. For accounting firms and corporate finance departments, this has profound implications. It offers a path to developing powerful analytical tools without incurring high operational costs or vendor lock-in. More importantly, a transparent model is inherently more auditable and explainable (XAI). System administrators and auditors can inspect the logic of the custom reranker, a critical feature when the AI's output is used to support high-stakes financial decisions or audit opinions.

Finally, the failure of our naive chunk expansion technique serves as a salient cautionary tale for AIS design and implementation. It empirically demonstrates the accounting principle of relevance over quantity. In a domain where materiality is paramount, simply flooding a model with more data is not only ineffective but detrimental, leading to higher rates of hallucination and less faithful answers. This finding reinforces the need for meticulous system design in accounting contexts. The goal must be to provide the decision-maker (whether human or AI) with concise, relevant, and high-quality information. This underscores that the successful application of AI in accounting depends less on brute-force data processing and more on the intelligent, structured curation of information, a core tenet of the Accounting Information Systems discipline.

\subsection{Limitations and Future Work}

While our proposed architecture proved highly effective, we acknowledge several limitations that pave the way for a rich agenda of future research at the intersection of AI and accounting information systems.

\begin{enumerate}
    \item Reliance on LLM-based Evaluation: The evaluation of answer quality was conducted using the RAGChecker framework. Future work should include a stage of manual evaluation by domain experts, such as chartered accountants or financial analysts, to provide a more nuanced assessment of answer quality and its fitness for professional use.
    \item Fixed Retrieval Parameters: Throughout our experiments, the number of retrieved chunks was fixed at a constant k=7 for all pipelines. This value was chosen as a reasonable baseline, but it is not necessarily optimal. Performance, particularly recall, might be improved by tuning this hyperparameter, and future work should investigate the impact of retrieving a larger or dynamically sized set of chunks.
    \item Refining Pre-Retrieval Intelligence: Our pre-retrieval steps showed mixed results. Future work could explore more sophisticated prompting strategies or fine-tune smaller LLMs on accounting-specific query transformation tasks to improve their reliability.
    \item Exploring Further Architectural Combinations: While this study evaluated a wide array of pipeline configurations, other combinations remain unexplored. For instance, the performance of our custom reranker without the pre-retrieval filtering step could be assessed, or different chunking strategies could be paired with our final architecture. A more exhaustive search of the architectural space could yield further incremental gains.
    \item Generalizability: This study was conducted on the FinanceBench dataset. While a robust benchmark, future work should validate the architecture's performance on a wider range of documents critical to accounting, not included in FinanceBench, such as sustainability (ESG) reports, proxy statements, and internal audit reports.
\end{enumerate}


\end{document}